\documentclass[aps,prl,superscriptaddress,twocolumn]{revtex4-1}

\usepackage{graphicx} 
\usepackage{amsmath}
 
\newcommand{\beq}{\begin{equation}}
\newcommand{\enq}{\end{equation}}
\newcommand{\bea}{\begin{eqnarray}}
\newcommand{\ena}{\end{eqnarray}}

\newcommand{\rr}{{\bf r}}

\begin{document}

\title{Vortex dynamics and skyrmions in four to six dimensions: Coherence vortices in Bose-Einstein condensates}
\author{Emil Lundh}
\affiliation{Department of Physics, Ume{\aa} University, 901 87 Ume\aa, Sweden}
\date{\today}
\begin{abstract}
I point out how coherence vortices, i.e., topological defects in a correlation 
function, could help explore new physics if they are created in matter waves. 
Vortex dynamics could be studied in up to six dimensions, and spin topological 
defects unseen in lower dimensions could be created. A rudimentary 
proof-of-principle experiment is sketched and simulated, in which three 
Bose-Einstein condensates are used to create and detect coherence vortices. 
\end{abstract}
\maketitle

Coherence vortices is a concept first coined in optics research \cite{gbur2003,gbur2005}.
These are phase singularities that exist not directly in a physical field, 
but in a spatial correlation function of a field. 
Typically, one may have a partially coherent field, such as a laser beam with 
limited coherence time, where the observed expectation value does not 
exhibit a phase singularity, but a double-slit interference experiment 
may reveal a phase winding around a 
singularity in the field-field correlation function. 
As such, coherence vortices are defined in a coordinate space with twice the 
dimensionality of the physical space. 
Simula and Paganin pointed out that this means that coherence vortices are 
possible in a one-dimensional system \cite{simula2011}; furthermore, they 
alluded to the possibility of observing coherence vortices in matter 
as well as in optics. 

A vortex in a two-point correlation function suggests the intriguing prospect 
of studying vortex dynamics in up to 
six dimensions. However, for the study of {\it dynamics}, optical systems are 
impractical and one needs to turn to coherent matter waves -- partially or 
fully Bose-Einstein condensed gases of ultracold atoms. 
Matter waves have several advantages as well as disadvantages with respect to 
optics when it comes to studying coherence vortices. Detection is a 
very different affair in matter waves, but it can be done, as 
will be discussed in detail below. Among the advantages are not only the 
possibility to study dynamics, but also the prospect of using multi-component 
condensates for creating higher-dimensional topological objects. 
A four- or six-dimensional space will offer new possibilities for these. 

The purpose of this paper is to point out the new prospects for research 
that are made possible by coherence vortices in matter systems; notably, 
high-dimensional vortex dynamics and high-dimensional topological defects. 
In addition, I sketch and simulate a clumsy but doable scheme for the 
realization and detection of coherence vortices in a system of trapped cold 
bosons. I see this as a proof of principle, which helps open up the way towards six-dimensional vortex dynamics. 

Let us first note that there exist already theoretical 
proposals that would in effect contain coherence vortices; namely the 
quantum fluctuating vortices studied by Martikainen and Stoof 
\cite{martikainen2004b}, and the slowly-rotating small Bose and Fermi systems 
studied by Reimann, Kavoulakis {\it et al.} \cite{kavoulakis2002,reimann2006}. 
What these systems have in common is that the exact ground state would 
contain a vortex whose position is subject to quantum uncertainty. 
The symmetry will be broken and a 
definite vortex position is chosen by any finite 
symmetry-breaking potential; this destroys the superposition and thus the 
coherence vortex is reduced to an ordinary vortex. 
The scheme I describe below is more robust, though perhaps less elegant.

{\it System.} -- 
Cold bosonic atoms are described by commuting field operators 
$\psi_a(\rr)$, $\psi_a^{\dagger}(\rr)$, where $a$ designates species and/or spin 
state ($F, m_F$), and $\rr$ is the spatial coordinate vector. 
If the atoms form a Bose-Einstein condensate (BEC), the expectation of a single field is non-vanishing, 
$\Psi_a(\rr) = \langle \psi_a(\rr)\rangle$,
and its expectation value is called the condensate wave function. 
The vortices supported by such wavefunctions are, of course, well 
studied \cite{fetter2009}.
Instead, we are here interested in phase singularities in the field-field 
correlation function, 
\beq
g_{aa}(\rr,\rr') = \langle \psi_a^{\dagger}(\rr) \psi_a(\rr')\rangle, 
\enq
which is a function of $2\times D$ coordinates, where $D$ is the number of 
dimensions in which the atoms can move. Such phase singularities may exist 
also when there are none in $\Psi_a(\rr)$. Henceforth, we will be dropping 
the subscript $a$ when we are dealing with only one component. 


{\it Topological defects.} -- 
In a single-component fluid in $D$ dimensions, 
the only type of topological defect that can 
exist is a vortex, characterized by an integer phase winding number around 
a closed loop. Such a loop surrounds a phase singularity of dimension $D-2$, 
i.e., a point vortex in two dimensions or a vortex line in three 
dimensions. In $D=4$ the singularity lies on a two-dimensional (2D) 
surface, and it is 
helpful to imagine the full 4D space as a succession of 3D spaces, each 
(except in degenerate cases) containing a vortex line, whose trajectory 
varies continuously as we move between adjacent 3D spaces. 
A closed loop is drawn around the vortex line in one of the 3D spaces. 
If the loop is displaced by a sufficiently small amount in any direction, 
including the fourth, it will still encircle the singularity. 
Thus the 2D sheet retains the topological stability we associate with 
a vortex. 

In a multi-component quantum fluid, higher-dimensional topological defects 
are possible, as is well known in the 3D case. 
The allowed topological defects in 
a given quantum fluid are classified using homotopy theory 
\cite{makela2003,yip2007,kibble2003}: Let $G$ be the 
symmetry group of the order parameter, and let $H$ be the group of operations 
that leave a specific ground state invariant. Then the order parameter space 
$M=G/H$ is is one-to-one correspondence with 
the set of distinct degenerate ground states of the system. Now 
the first homotopy group of $M$, $\pi_1(M)$, gives the allowed topological 
defects that may be encircled by a ring; these are (singular) 
point defects in 2D, line 
defects in 3D, and planes in 4D. 
Likewise, the second homotopy group 
$\pi_2(M)$ enumerates those topological defects that can be encircled by 
a spherical shell, and in 3D those are point defects, i.e., monopoles. 
For instance, in a single-component BEC, $M=S^1$, the circle; and 
we have $\pi_1(S^1)=Z$, where $Z$ is the set of integers. These integers 
are the possible winding numbers for a vortex. 
It is now plain to see that in higher dimensions, 
there may be topological defects classified 
by higher-order homotopy groups, which do not make sense 
in lower dimensions -- e.g., $\pi_3$ characterizes 
singular point defects in 4D or sheet defects in 6D. 
In a single-component BEC, again, all homotopy groups higher than 1 are 
zero; $\pi_2(S^1)=\pi_3(S^1)=0$ etc; 
which implies that the only topologically stable 
defect for such a fluid is the vortex. 
For the ferromagnetic state of a spin-1 condensate, $M$=SO(3), 
and its homotopy groups are $\pi_3(M) = Z$ and $\pi_4(M)=\pi_5(M)=Z_2$, 
where $Z_2$ is the set of integers modulo 2.
Further investigation of the homotopy groups associated with the 
correlation function matrices $g_{ab}$ of the ground states in 
spinor condensates must be left to future studies.

{\it Creation. -- } 
We now turn to the experimental creation and detection of coherence 
vortices. 
A number of ingredients are necessary for the creation, and a few more for the 
detection, of coherence vortices. First, one needs a non-trivial correlation 
function. 
If the whole system is fully coherent -- i.e., one pure BEC -- 
vortices exist in the correlation function if and only if they 
are there in the condensate wavefunction itself, 
and the study of the associated coherence vortices would be pointless. 
A partially coherent cloud, such as bosons in a one-dimensional or disordered 
potential, will also not present coherence vortices in most cases, but 
only a decaying correlation function. 
Instead, I sketch an experiment 
that uses the random relative phase of two independent BECs. 

The experiment starts out with a pair of independent BECs, for simplicity 
in two dimensions. 
Pairs of independent BECs are created by holding atoms in two separate traps 
while cooling them down below the critical temperature \cite{andrews1997}. 
Such pairs of BECs will produce interference patterns when brought 
together, but the two are statistically independent in the sense that 
their global phase difference -- and thus the phase of the interference 
pattern -- varies randomly from shot to shot. Interference between the 
two independent BECs is, thus, washed out only in the average over many 
experimental runs. 
In an optical system, a beam has a finite coherence time and thus it 
self-averages. 
It is possible that one could construct something similar by making clever 
use of the finite coherence length of an elongated or disordered system, but I 
will not follow that line of thought here. 
I thus opt for averaging over repeated single shots as the straightforward 
(but tedious) way to construct an ensemble average -- by hand, so to speak -- 
in this proposed experiment. 

The BECs are put in different spin states and 
released towards each other head on. 
The two spin states $a$ will be labeled 1 and 2, respectively, and the 
trapping potential for each state is written as 
\beq
V_a(\rr) = \frac12 m\omega^2(\rr-\rr_{0a}),
\enq
where the trap center $\rr_{0a}$ is taken to be different for the two 
BECs. They are first Bose-Einstein condensed at a distance 
large enough to avoid contact, and then adiabatically brought to 
their starting positions, $\rr_{0a}$. 
At time $t=0$ the traps are suddenly 
moved to target positions $\rr'_{0a}$. 
The BECs are accelerated towards these new potential minima, and 
the resulting collision excites vorticity. 
At a final time $t$, the atoms of species 1 are transferred into species 2, 
and the resulting state is recorded. 
Direct observation of the density profile will reveal a set of vortices, 
whose positions depend on the initial, random phase difference between 
the two BECs. We denote this phase difference $\Theta$. 
It is in the transfer at time $t$ that this phase difference will 
be decisive. 

\begin{figure}
\includegraphics[width=0.45\textwidth]{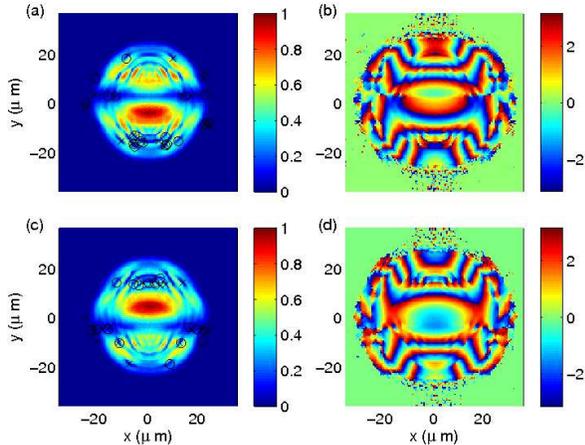} 
\caption{[Color online] 
Collision of two BECs for two cases 
with different and randomly chosen initial phase difference. 
Simulation parameters as in text. 
Density and phase are recorded at $t=100$~ms. 
(a), (c): Density; (b), (d): Phase. 
(a), (b): $\Theta=1.98296$ radians; (c), (d): $\Theta=4.19545$ radians.
Units for the density is such that its maximum value is 1.
Positions of vortices are indicated with crosses in the figures, 
and antivortices with rings. 
}
\label{fig:singleshots}
\end{figure}

\begin{figure}
\includegraphics[width=0.21\textwidth]{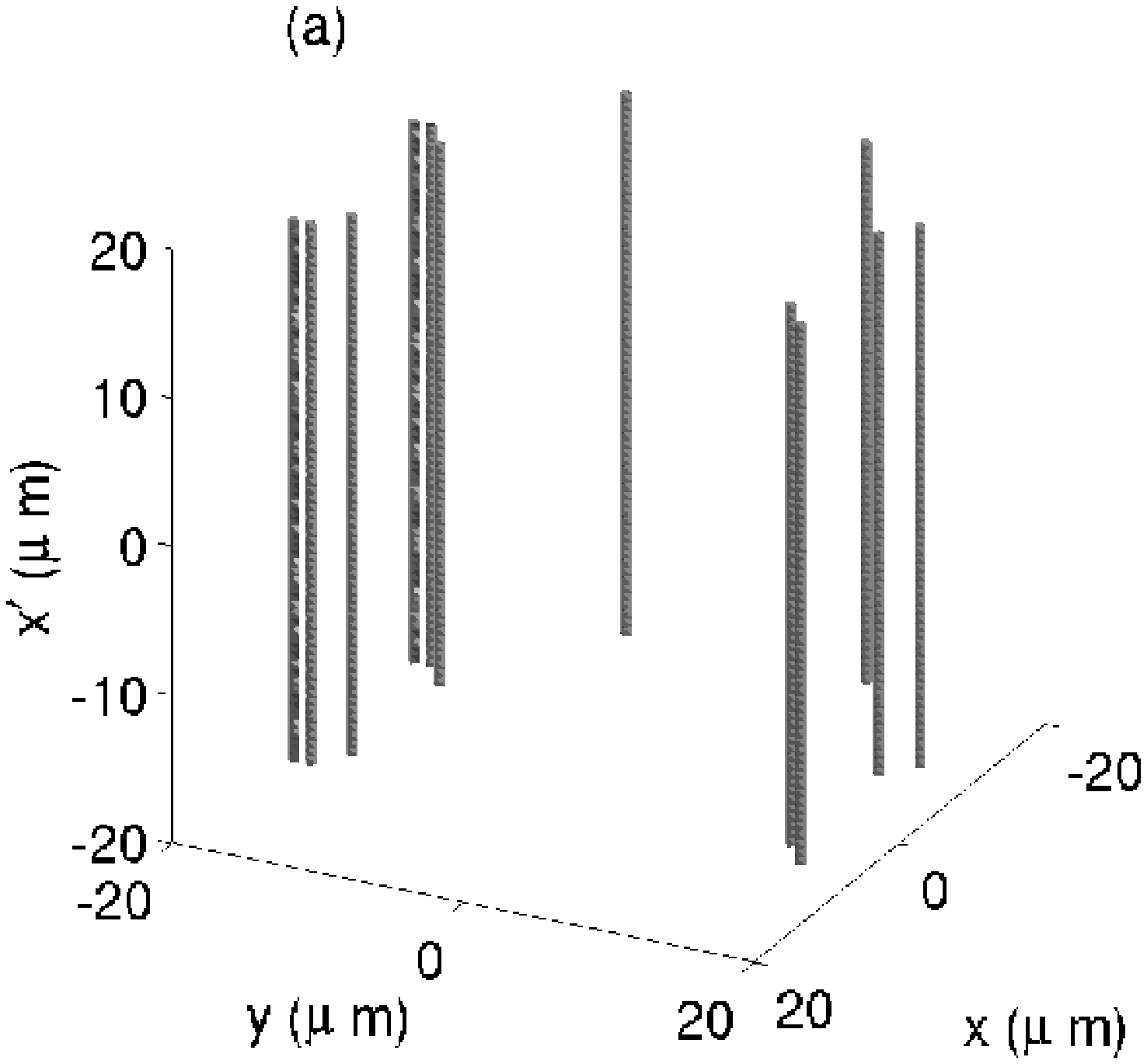} 
\includegraphics[width=0.21\textwidth]{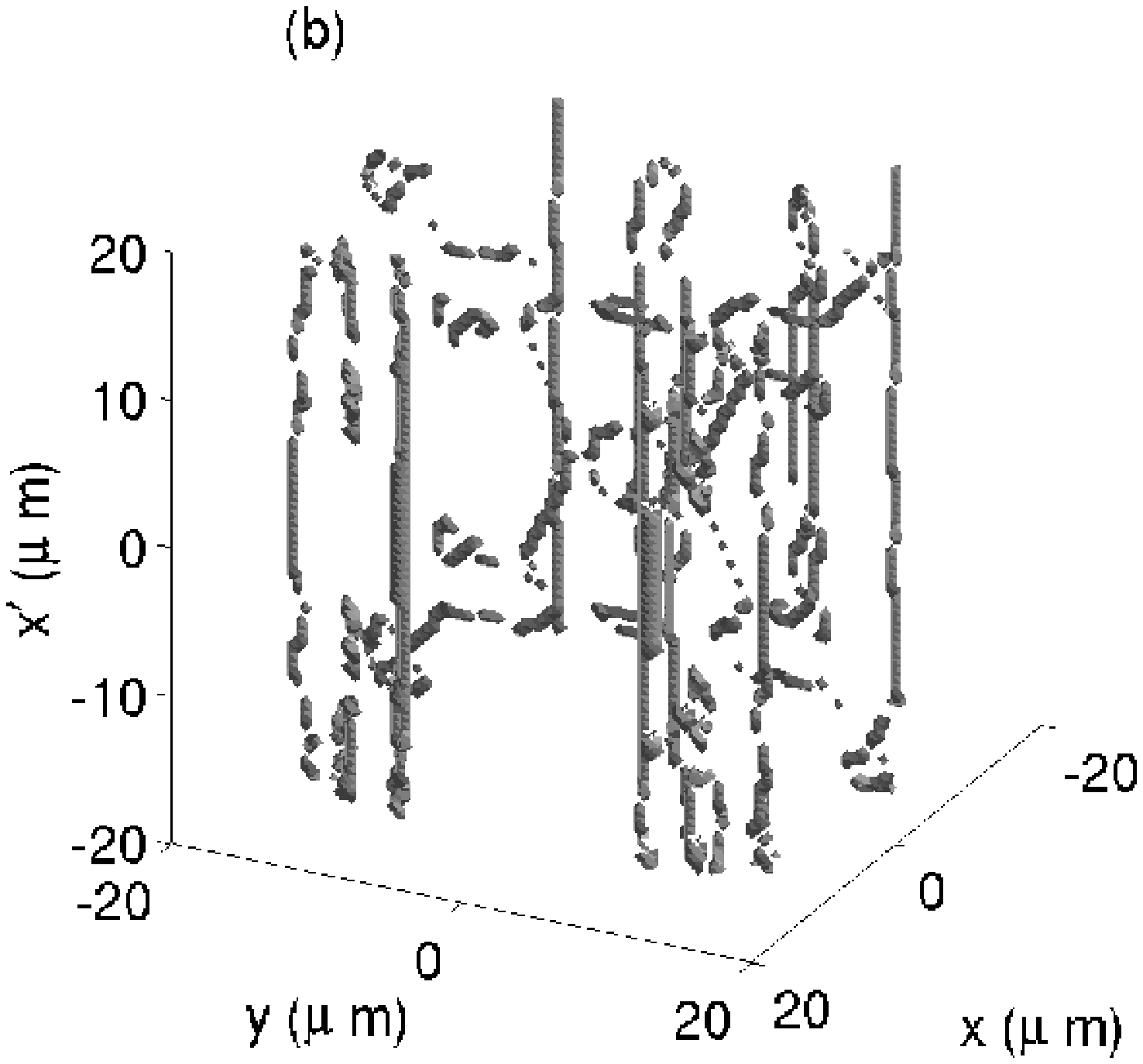} 
\includegraphics[width=0.21\textwidth]{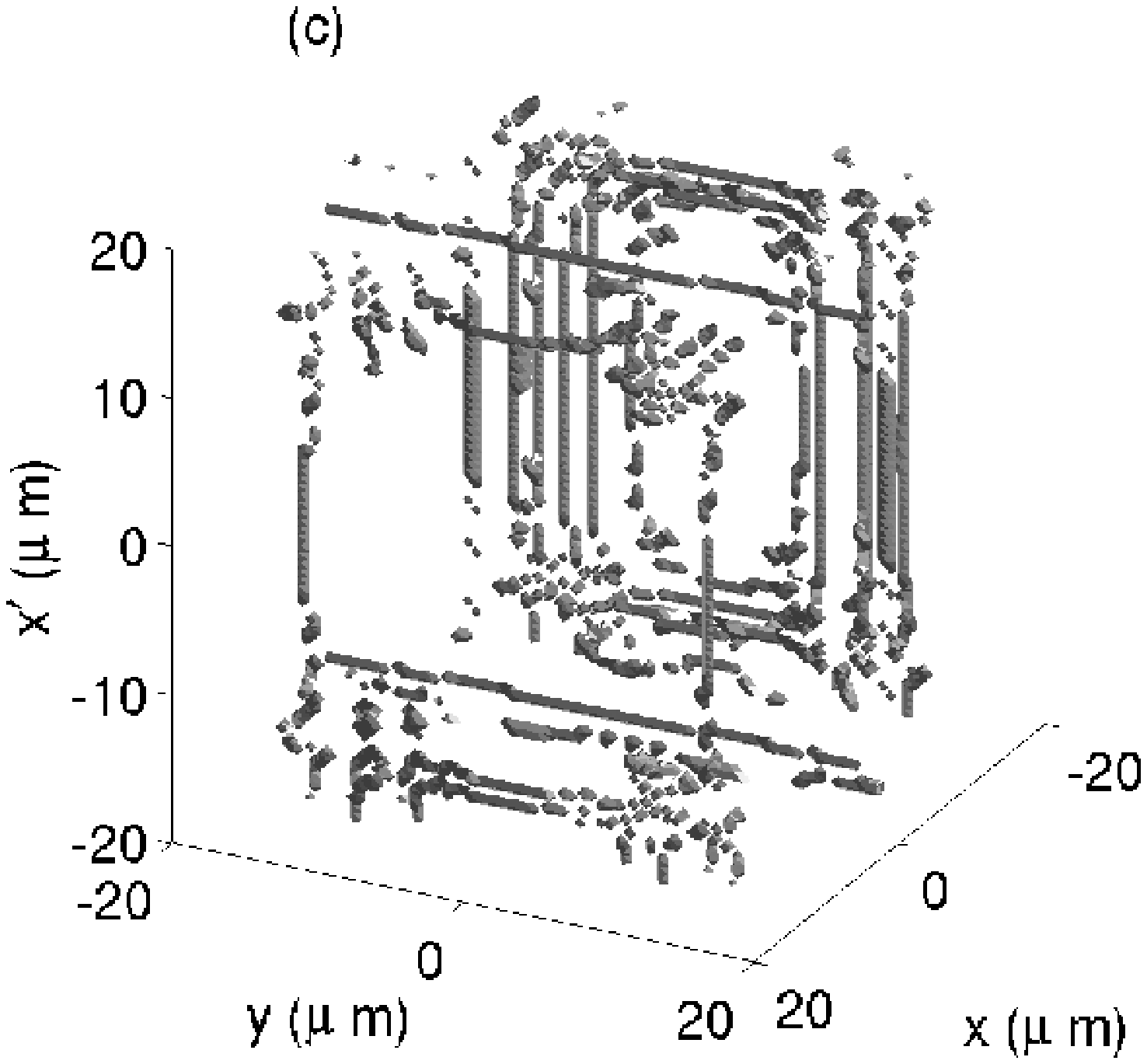} 
\includegraphics[width=0.21\textwidth]{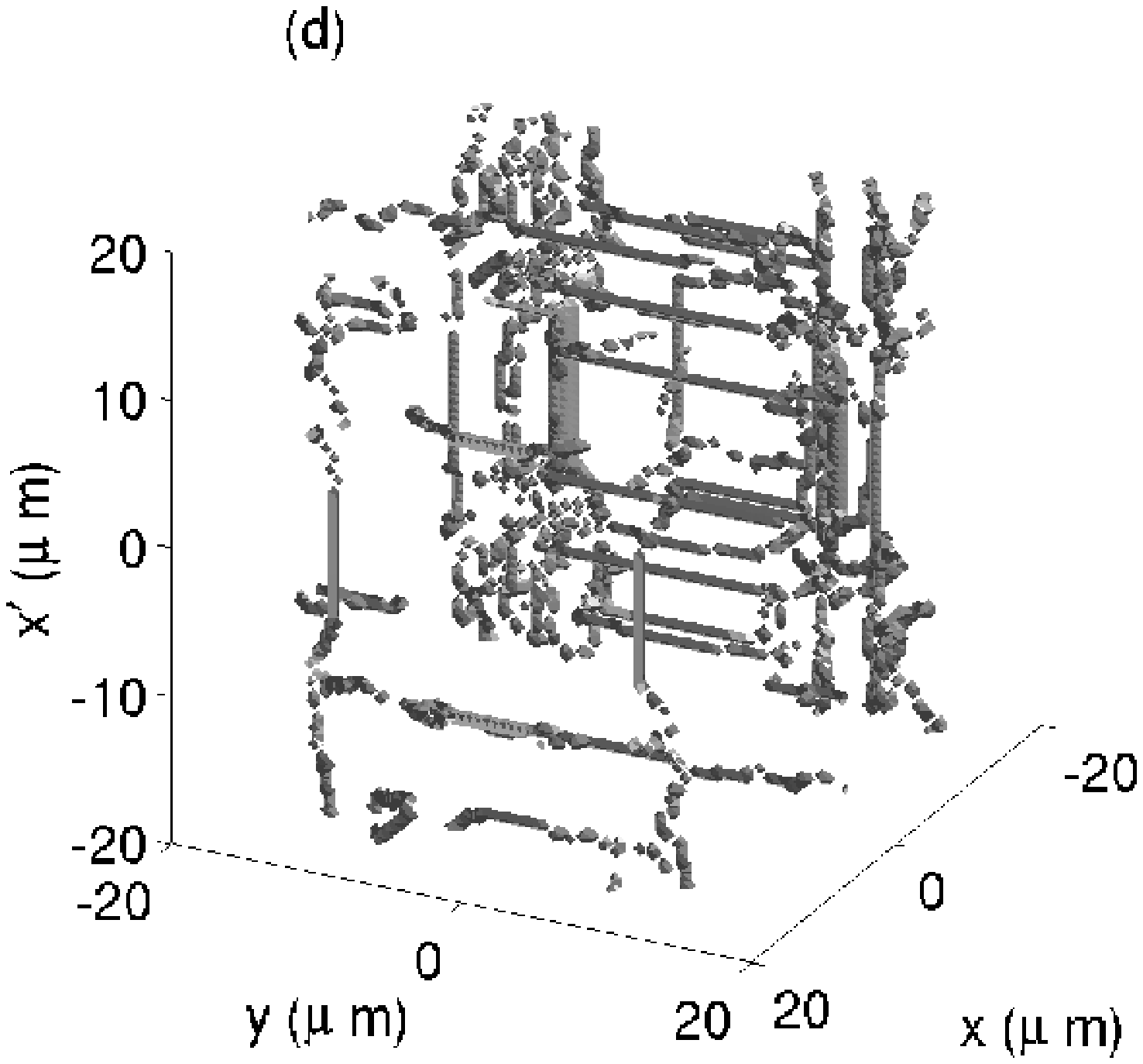} 
\caption{
Vortices (isosurfaces of the phase winding) 
of the correlation function 
$g(x,y,x',y')$, at three-dimensional slices of the four-dimensional 
space, fixed at 
(a) $y'=0$, (b) $y'=3.42 \mu$m, (c) $y'=6.83 \mu$m, and (d) $y'= 10.25$, respectively.
The parameters are as in Fig.\ \ref{fig:singleshots}.
}
\label{fig:fourdee}
\end{figure}

The process is simulated for a binary mixture of $4.4\times 10^{4}$ atoms of 
$^{87}$Rb equally distributed among the states $|F, m_F\rangle=|1,1\rangle$ and 
$|F,m_F\rangle=|1,-1\rangle$, each in a trap of frequency 
$\omega=2\pi\cdot 20$~s$^{-1}$ in the plane, 
and in addition a tight trap of frequency 
$2\pi\cdot 400$~s$^{-1}$ is assumed along the $z$ direction; this direction is 
integrated out so that the physics in the 2D plane is simulated.
The starting positions are at $\rr_{01}=(12~\mu{\mathrm m}, 0)$, and $\rr_{02}=-\rr_{01}$, 
and the target positions are both at the origin, $\rr'_{01}=\rr'_{02}={\bf 0}$.  
This system is described by the 
two-component Gross-Pitaevskii equation, 
\begin{eqnarray}
\label{GP1}
i \hbar\frac{\partial }{\partial t}{{\Psi}_{1}}=\left[ -\frac{{{\hbar }^{2}}}{2{{m}}}\nabla^2+{{{V}}_{1}}+{{U}_{11}}{{\left| {{{\Psi}}_{1}} \right|}^{2}}+{{U}_{12}}{{\left| {{{\Psi}}_{2}} \right|}^{2}} \right]{{{\Psi} }_{1}}, \\
\label{GP2}
i \hbar\frac{\partial }{\partial {t}}{{{\Psi} }_{2}}=\left[ -\frac{{{\hbar }^{2}}}{2{{m}}}\nabla^2+{{{V}}_{2}}+{{U}_{22}}{{\left| {{{\Psi}}_{2}} \right|}^{2}}+{{U}_{12}}{{\left| {{{\Psi} }_{1}} \right|}^{2}} \right]{{{\Psi}}_{2}},
\end{eqnarray}
with $U_{ab}=4\pi\hbar^2a_{ab}/m$, where the $s$-wave scattering lengths are 
$a_{11}=a_{22}=100.4a_{\rm B}$ and $a_{12}=101.3a_{\rm B}$, where $a_{\rm B}$ is 
the Bohr radius. 
A definite value of $\Theta$ is used in each simulation run. 
Two examples, for different $\Theta$, are shown in Fig.\ \ref{fig:singleshots}.
%
%
%
%
The physics of the violent collision is seen to be mainly governed 
by pass-through \cite{sasaki2009} and 
quantum interpenetration \cite{kobyakov2012a}. 
Subsequent transfer of the population to component 1 results 
in a single BEC containing a set of vortex-antivortex pairs whose positions -- 
in the Gross-Pitaevskii picture -- depend on $\Theta$. 
Averaging over the phase difference will wash out 
the vortices. 
However, the vorticity in the correlation function will not. 
The exact correlation function after transfer is simply 
\beq
g(\rr,\rr') = \Psi_1^*(\rr)\Psi_1(\rr')+\Psi_2^*(\rr)\Psi_2(\rr'),
\enq
since the cross-term vanishes for independent condensates.
The vortices in 
$g(x,y,x',y')$ 
corresponding to Fig.\ \ref{fig:singleshots} 
are imaged in Fig.\ \ref{fig:fourdee} as a sequence of 3D frames 
in the space $(x,y,x')$, with a fixed $y'$ for each frame. 
There are vertical vortex lines visible in all four isosurfaces; 
these are inherited from the vorticity created in the 
collision (cf.\ \cite{gbur2005}). In addition, a tangle is created of 
vortices not perpendicular to the $xy$ plane; 
it is tempting to think of these as more genuine coherence vortices. 

Note that before the transfer, the vorticity in the two-component 
system takes the form of skyrmions or filled-core vortices. 
However, it would be incorrect to call them coherence vortices, since 
their locations are independent of the phase difference $\Theta$ and 
thus they are manifest directly in the density itself, even after 
averaging.

{\it Detection.} --  
In order to observe the coherence vortices experimentally, 
one must reconstruct the 
field-field correlation function $g(\rr,\rr')$. 
For cold atoms, there exists no practical method for measuring it
at arbitrary coordinates. 
For optical systems, one may let the light pass 
through two pinholes and observe the phase of the resulting interference 
pattern. Ref.\ \cite{simula2011} mentions this type of method as an 
alternative also for matter systems, but does not discuss how. 
It is possible that a scheme could be devised where atoms 
in small neighborhoods of two points are excited and then brought together. 
Instead, one may obtain an approximate correlation function by 
experimentally imitating the type of averaging 
that is done in c-field calculations \cite{bezett2008}. 

\begin{figure}[htbp]
\includegraphics[width=0.45\textwidth]{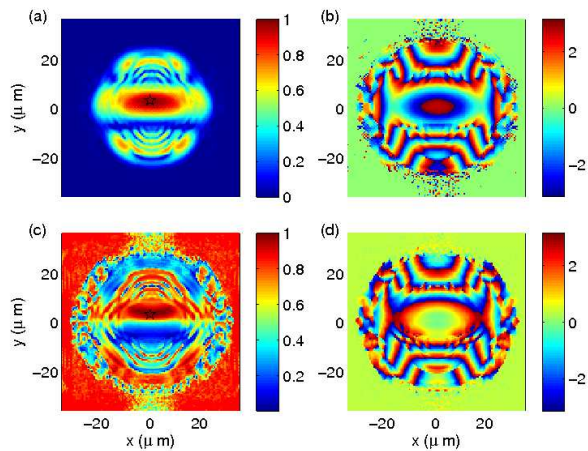} 
\caption{[Color online] 
Field-field correlation function $g(\rr,\rr'_0)$, with $\rr_0$ 
fixed. The position of $\rr_0$ is indicated with a 
star. 
(a), (b): Exact correlation function. 
(c), (d): Approximate correlation function of Eq.\ (\ref{expcorrelation}), 
which simulates the result of the proposed experiment. 
(b), (d): Phase of $g$; (b), (d): Absolute value $|g|$. 
Normalization is such that the maximum value of $|g|$ is 1.
The parameters are as in Fig.\ \ref{fig:singleshots}.
}
\label{fig:comparewithdens}
\end{figure}

In a single realization, the spatial profile of the condensate's phase 
can be measured by letting the BEC 
interfere with another one initially held at a distance. 
(In the present scheme, this means managing three BECs, whereof 
two should be independent.) 
One can now record the set of phase differences between all points in the 
single run, and then average this function over repeated runs to obtain 
the approximate correlation function, 
\beq
g_{\rm exp}(\rr,\rr') = \langle e^{i[\theta(\rr')-\theta(\rr)]} \rangle,
\label{expcorrelation}
\enq
where $\theta(\rr)$ is the phase at point $\rr$ inferred from the 
fringe pattern in a single experimental run. 
The approximation built into this procedure is the neglect of spatial 
density variations, as seen in Eq.\ (\ref{expcorrelation}). 
It is often a good one as far as the locations 
of the centermost vortices are concerned. 
We simulate an actual experiment by making 30 individual runs with 
random initial values of $\Theta$ and random noise on the wavefunctions 
$\Psi_j$ on a 5\% level. 
In Fig.\ 
\ref{fig:comparewithdens}, the simulated correlation function $g_{\rm exp}$ is 
compared with the exact one that includes density information. 
%
%
The result may be improved by reading off a coarse-grained 
density profile from the interference fringes, if one so wishes. 

The time development of the coherence vortices may now be studied by 
doing repeated experiments; since we are studying average quantities, 
the destructive imaging technique is not a problem. 

\begin{figure}[htbp]
\includegraphics[width=0.45\textwidth]{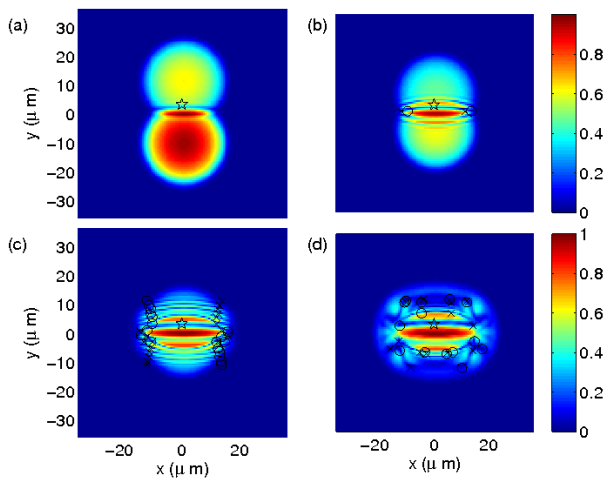} 
\caption{[Color online] 
Time development of the field-field correlation function 
$g(\rr,\rr'_0)$, with $\rr_0$ fixed. 
The modulus $|g(\rr,\rr_0')|$ is displayed. 
(a): Time $t=25$ms, (b): $t=50$ms, (c): $t=75$ms, (d): $t=100$ms. 
Positions of coherence vortices are indicated with crosses in the figures, 
and antivortices with rings. The position of $\rr_0$ is indicated with a 
star. 
The parameters are as in Fig.\ \ref{fig:singleshots}.
}
\label{fig:timedevelop}
\end{figure}

Correlation functions for four time instances are shown in Fig.\ \ref{fig:timedevelop}.
It can be seen how the behavior of the correlation function mirrors 
the physical process: two condensates approaching, colliding and 
interpenetrating, creating vortex-antivortex pairs in the process. 
This picture is made especially clear by choosing the 2D plane of view 
to be parallel to the physical space; as we have seen, vortex tangles 
move in all four dimensions.

{\it Concluding remarks.} -- Summing up, this Letter wishes to point out 
two (related) reasons for studying coherence vortices in cold matter systems: 
It opens up for studying dynamics of vortices in high dimensions, and 
new types of high-dimensional topological defect, unseen in 3D, can be 
created here. 
As a specific example, a system of two colliding BECs 
was simulated, in which a tangle of two-dimensional coherence vortex sheets 
was created. A recipe for a proof-of-principle experiment was laid out 
where coherence vortices are created and detected using current 
experimental technology. 

The author is indebted to Harri M\"akel\"a for help with homotopy groups and 
to Jani Martikainen for feedback on the manuscript.


\begin{thebibliography}{10}

\bibitem{gbur2003}
G. Gbur and T. Visser, Opt. Comm. {\bf 222},  117  (2003).

\bibitem{gbur2005}
G. Gbur and T. Visser, Opt. Comm. {\bf 259},  428  (2005).

\bibitem{simula2011}
T.~P. Simula and D.~M. Paganin, Phys. Rev. A {\bf 84},  052104  (2011).

\bibitem{martikainen2004b}
J.-P. Martikainen and H.~T.~C. Stoof, Phys. Rev. A {\bf 70},  013604  (2004).

\bibitem{kavoulakis2002}
G.~M. Kavoulakis, S.~M. Reimann, and B. Mottelson, Phys. Rev. Lett. {\bf 89},
  079403  (2002).

\bibitem{reimann2006}
S.~M. Reimann, M. Koskinen, Y. Yu, and M. Manninen, Phys. Rev. A {\bf 74},
  043603  (2006).

\bibitem{fetter2009}
A.~L. Fetter, Rev. Mod. Phys. {\bf 81},  647  (2009).

\bibitem{makela2003}
H. M\"akel\"a, Y. Zhang, and K.-A. Suominen, J. Phys. A: Math. Gen. {\bf 36},
  8555  (2003).

\bibitem{yip2007}
S.-K. Yip, Phys. Rev. A {\bf 75},  023625  (2007).

\bibitem{kibble2003}
T. Kibble,  in {\em Conference of the NATO-Advanced-Study-Institute on Patterns
  of Symmetry Breaking}, edited by H. Arodz, J. Dziarmaga, and W.~H. Zurek
  (Kluwer, Dordrecht, 2003), pp.\ 3--36.

\bibitem{andrews1997}
M.~R. Andrews {\it et~al.}, Science {\bf 275},  637  (1997).

\bibitem{sasaki2009}
K. Sasaki, N. Suzuki, D. Akamatsu, and H. Saito, Phys. Rev. A {\bf 80},  063611
   (2009).

\bibitem{kobyakov2012a}
D. Kobyakov {\it et~al.}, Phys. Rev. A {\bf 85},  013630  (2012).

\bibitem{bezett2008}
A. Bezett, E. Toth, and P.~B. Blakie, Phys. Rev. A {\bf 77},  023602  (2008).

\end{thebibliography}

\end{document}